\documentclass{elsart}

\usepackage{graphicx}

\begin{document}               

\def\be{\begin{equation}}
\def\ee{\end{equation}}
\def\ba{\begin{eqnarray}}
\def\ea{\end{eqnarray}}
\def\bas{\begin{eqnarray*}}
\def\eas{\end{eqnarray*}}

%\addtolength{\topmargin}{2cm}
\begin{frontmatter}
\title{Origin of chaos in the spherical nuclear shell model: role of
symmetries}
\author[ut,ornl]{T.~Papenbrock}
\and
\author[mpi]{H.~A.~Weidenm\"uller}
\address[ut]{Department of Physics and Astronomy, University of Tennessee,
Knoxville,~TN~37996, USA}
\address[ornl]{Physics Division, 
Oak Ridge National Laboratory, Oak Ridge, TN 37831, USA}
\address[mpi]{Max-Planck-Institut f\"ur Kernphysik, D-69029 Heidelberg,
Germany}
\date{\today}

\begin{abstract}
To elucidate the mechanism by which chaos is generated in the shell
model, we compare three random--matrix ensembles: the Gaussian
orthogonal ensemble, French's two--body embedded ensemble, and the
two--body random ensemble (TBRE) of the shell model. Of these, the last
two take account of the two--body nature of the residual interaction,
and only the last, of the existence of conserved quantum numbers like
spin, isospin, and parity. While the number of independent random
variables decreases drastically as we follow this sequence, the
complexity of the (fixed) matrices which support the random variables,
increases even more. In that sense we can say that in the TBRE, chaos
is largely due to the existence of (an incomplete set of) symmetries.

\end{abstract}
\end{frontmatter}

\section{Introduction and Motivation} 

The analysis of nuclear spectra has produced ample evidence for
chaotic motion. Indeed, near neutron threshold, the spectra of
medium--weight and heavy nuclei display fluctuations which agree with
those of random matrices drawn from the Gaussian orthogonal ensemble
(GOE)~\cite{HaPaBo}. Similar agreement has been found for nuclei in
the $sd$--shell (both in experimental data~\cite{Mitchell} and in
shell--model calculations~\cite{Zelevinsky}), and in the ground--state
domain of heavier nuclei~\cite{Harney}, although here there exists
strong evidence, too, for regular motion as predicted by the shell
model and the collective models. Calculations in Ce~\cite{Flambaum}
have produced similar evidence for chaotic motion in atoms. Thus,
chaos appears to be an ubiquitous feature of interacting many--body
systems. What is the origin of this behavior? In the present paper,
we address aspects of this question.

We do so using the nuclear shell model, a theory with a mean field and
a residual two--body effective interaction $V$. (We do not include
three--body forces, although there is evidence~\cite{threebody} that
these may be needed to attain quantitative agreement with data. It will
be seen that qualitatively, our arguments would not change with the
inclusion of such forces.) In many nuclei, the mean field is (nearly)
spherically symmetric. Thus, single--particle motion is largely
regular. Chaos in nuclei seems a generic property and, hence, must be
due to $V$. We focus attention entirely upon the effects of $V$.
Therefore, we assume that we deal with a single major shell in which
the single--particle states are completely degenerate and in which
there is a fixed number of valence nucleons. (A lack of complete
degeneracy would reduce the mixing of states due to $V$ and, thus,
drive the system towards regular motion). Generic results are expected
to be independent of the details of $V$. Therefore, we assume that the
two--body matrix elements (TBME) of $V$ are uncorrelated
Gaussian--distributed random variables with zero mean value and unit
variance. Our results then apply to almost all two--body interactions
with the exception of a set of measure zero. (The integration measure
is the volume element in the parameter space of the TBME.) The
resulting random--matrix model is commonly referred to as the two--body
random ensemble (TBRE)~\cite{French,Bohigas}. We ask: How does $V$
produce chaos in the framework of the TBRE?

The two--body interaction $V$ has two characteristic features. (i)
It connects pairs of nucleons. (ii) It possesses symmetries: it
conserves spin, isospin, and parity. We wish to elucidate the role of
both features in producing chaos in nuclei.

The relevance of the first feature is brought out by comparing the
TBRE with the GOE. We recall that in the latter, the matrix elements of
the Hamiltonian couple every state in Hilbert space to every other such
state. These matrix elements are assumed to be uncorrelated random
variables. In the context of many--body theory, such independent
couplings between all pairs of states can be realized only in terms of a
many--body interaction the rank of which equals the number of valence
particles. Put differently, with $N$ the dimension of the Hamiltonian
matrix, the number of independent random variables in the GOE is
$N(N+1)/2$ and, for $N \to \infty$, grows much faster than $N$.
Thus, it is intuitively clear that the GOE Hamiltonian will produce a
thorough mixing of the basis states which is tantamount to chaos. In
contradistinction, the number of independent two--body matrix elements
in a single shell with half--integer spin $j$ is only $j + 1/2$ while
the number of many--body states with fixed total spin $J$ grows with $j$
like $j^{m - 3}$ where $m$ is the number of valence particles. (The
simple estimates leading to these statements are given in the Appendix.
The statements apply for $j \gg m \gg 1$, and $J \ll m$). Thus,
in the TBRE the number of independent random variables is much smaller
than the dimension of typical matrix spaces, and it is a non--trivial
fact that $V$ produces as much mixing of the basis states as the GOE
Hamiltonian. We wish to elucidate the mechanism which is responsible
for this mixing.

As for the second feature (the role of symmetries), we compare the TBRE
with another random--matrix model which lacks the symmetries of the TBRE
but likewise assumes a random two--body interaction. This is the embedded
two--body ensemble of Gaussian orthogonal random matrices
(EGOE(2))~\cite{MF}. (For a recent review we refer the reader to
Ref.~\cite{Kota}). In this model, $m$ fermions are distributed over
$l > m$ degenerate single--particle states. Hilbert space is spanned by
the resulting $N = {l \choose m}$ Slater determinants. The two--body
interaction connects only those Slater determinants which differ in the
occupation numbers of not more than two single--particle states.
Therefore, the representation of the two--body interaction in the Hilbert
space of Slater determinants yields a sparse matrix (most non--diagonal
matrix elements vanish). This model does not respect the symmetries of
the shell model. Neither the single--particle states nor the two--body
interaction carry any quantum numbers. Obviously, the model is very
different from the GOE. It is likewise very different from the TBRE.
In the latter, the single--particle states do carry spin, isospin, and
parity quantum numbers, and $V$ conserves these symmetries. Total Hilbert
space decays into orthogonal subspaces carrying these same quantum
numbers. Each subspace is spanned by states which are linear combinations
of (many) Slater determinants. As a result, the matrix representation of
$V$ in any such subspace becomes fairly dense, even though it remains
true that $V$ connects only Slater determinants which differ in the
occupation numbers of not more than two single--particle states.

In comparing the EGOE(2) and the TBRE, one may consider several options.
(i) One might use the fixed Hilbert space of all many--body states
that exist within a given major shell, with all possible quantum
numbers for total spin, parity, and isospin. These states are coupled
either via a symmetry--preserving random interaction (this is the
TBRE; here the Hamiltonian has block--diagonal structure), or via a
symmetry--breaking random interaction (this is the EGOE(2)). One might
compare the spectral statistics of the eigenvalues and the mixing of the
eigenfunctions in both models. At this point in time, such a comparison
is impractical because of the huge dimension of the matrices involved
for the EGOE(2). (ii) One might use a Hilbert space of many--body states
with fixed values for total spin, parity and isospin. These states are
coupled again either via a symmetry--preserving random interaction
(this is the TBRE within the subspace of many--body states with fixed
quantum numbers) or, in the case of a symmetry--breaking interaction,
via effective interactions that take into account the coupling to
many--body states with different quantum numbers. Here the construction
of the effective interaction poses severe difficulties and is
practically impossible for realistic cases. Both options (i) and (ii)
ask questions relating to the physical role of conserved quantum
numbers in the shell model. Even if these options were open, we would
probably not have used them. The aim of this paper is different. We
wish to elucidate the mechanisms which are operative in different
random--matrix models (GOE, EGOE(2), TBRE) in generating chaos. This is
why we have followed option (iii): The role of symmetries can also be
displayed by comparing the structure of the Hamiltonian matrices which
are typical for the TBRE and the EGOE(2), in both cases for Hilbert
spaces of large dimensions. This option does not require the
diagonalization of huge matrices or the difficult construction of
effective interactions. It dodges the issue of the physical role of
symmetry in shell--model calculations and focusses instead upon the
structural aspects of the matrices involved in the two approaches.

In Section~\ref{abs} we elucidate the structural features of the
EGOE(2). In Section~\ref{pres} we then turn to a corresponding analysis
for the TBRE in the case of a single $j$--shell, and of the
$sd$--shell. For both models we aim at displaying those structural
elements which are absent in the GOE and which are essential for
producing a thorough mixing of the basis states in Hilbert space and,
thus, chaos, in spite of a strongly reduced number of independent
random variables. The comparison between the EGOE(2) and the TBRE will
then display the role of symmetries in the TBRE. (This aspect of the
TBRE has been coined ``geometric chaoticity'' by Zelevinsky {\it et
al.}~\cite{Zelevinsky}. To the best of our knowledge, however, the
actual role of symmetries in the TBRE has never been investigated.) We
are led to the conclusion, that symmetries are a vital element of the
TBRE and, in this sense, instrumental in producing chaos in nuclei.

We are aware of a huge body of literature addressing issues closely
related to our theme. Some of these are reviewed in
Refs.~\cite{Zelevinsky,Kota,BW03}. As mentioned above, there is ample
evidence for chaos in nuclear shell--model calculations in the
$sd$--shell and beyond. Likewise, there is evidence for chaos in the
EGOE(2), at least in the center of the spectrum. Even deviations from
complete chaos have been understood since a long time. As early as 1979,
it was, f.i., shown~\cite{ver} that the non--degeneracy of the
single--particle states in the $sd$--shell prevents a complete mixing
of all the states of given total spin and causes the partial width
amplitudes (given as projections of the eigenstates of the shell--model
Hamiltonian onto a fixed vector in Hilbert space) to deviate from the
Porter--Thomas distribution predicted by the GOE. We are not concerned
with adding to this impressive body of evidence. We take it for granted
that chaos exists in the shell model and in the EGOE(2). Rather, we
wish to understand the structural aspects of the TBRE and of the EGOE
which produce chaos.

We believe that our investigation offers novel insights into the origin
of chaos in nuclei in the following two respects: (i) We demonstrate
that chaos is generic and ubiquitous both in a single $j$--shell and in
the $sd$--shell. To the best of our knowledge, previous arguments have
always relied upon numerical results based upon a {\it specific} choice
of the two--body interaction. In contradistinction, we use {\it generic}
aspects of the TBRE to show that we must always expect strong mixing of
the basis states. (ii) We exhibit the role played by symmetries in the
TBRE for the generation of chaos by comparing this ensemble with the
EGOE(2), and with the GOE. We are not aware of any previous work on this
topic.

Our emphasis on the role of symmetry in generating chaos may be
surprising. In fact, the existence of a complete set of quantum
numbers (usually connected to symmetries of the Hamiltonian) is
equivalent to integrability and, thus, diametrically opposed to chaos.
The case of nuclei (and, for that matter, of atoms) is different. The
symmetries that dominate nuclei (invariance under rotation, mirror
reflection, and proton--neutron exchange) are incomplete (they do not
form a complete set of integrals of the motion). Taking account of
these symmetries, we can write the Hamiltonian in block--diagonal
form. Chaotic motion does seem to exist in every such block. Therefore,
it is meaningful to ask: In which way is the origin of chaos influenced
by the presence such an incomplete set of quantum numbers? It is this
question which we answer by comparing the TBRE with the EGOE(2) and the
GOE.

\section{Absence of Symmetries} 
\label{abs}

In this Section, we analyse the EGOE(2) and then compare it with the
GOE. In the absence of symmetries, the appropriate random two--body
interaction model is the EGOE(2). The EGOE(2) is often used to model
stochastic aspects of realistic systems like small metallic grains or
quantum dots. This is justified since in those systems the
single--particle wave functions themselves are chaotic, and the
resulting two--body matrix elements reflect this property and transport
the information of the underlying one--body chaos into the many--body
system. Several numerical studies for matrix dimensions up to a few
thousand or so have shown that the EGOE(2) exhibits GOE statistics in
the center of the spectrum. For infinite matrix dimensions, the
situation is less clear. Evidence from previous work on the
EGOE(2)~\cite{BRW01,BW03} suggests that the level statistics is
Poissonian. However, no firm conclusion has yet been
reached~\cite{Srednicki}. Here, we supplement the previous analysis by
a different approach and focus on the matrix structure. In the
EGOE(2), the random variables are the $a = { l \choose 2} 
[{l \choose 2} + 1]/2$ independent two--body matrix elements
$V_\alpha, \alpha = 1, \ldots, a$. A Hamiltonian drawn from the
EGOE(2) has matrix elements
\be
\label{HamEGOE}
H_{\mu \nu}=\sum_{\alpha=1}^a V_\alpha D_{\mu \nu}(\alpha) \ , 
\ee 
where the matrices $D_{\mu \nu}(\alpha)$ transport the information
contained in the two--body matrix elements $V_\alpha$ into the
$N$--dimensional Hilbert space spanned by Slater determinants labeled
$\mu$ or $\nu$.

The matrices $D$ play a central role in the understanding of the
EGOE(2). Indeed, these matrices are the structural elements of this
ensemble, while the $V_\alpha$'s are just a set of random variables
that change from realization to realization. Pictorially speaking,
the matrices $D$ form the scaffolding which supports the random
variables $V_\alpha$. The properties of the EGOE(2) (averages, higher
moments and correlation functions) of both the Hamiltonian and the
Green's functions are completely determined by the matrices $D$.

The ensemble average of the Hamiltonian is obviously zero. The second
moment is 
\be
\label{HamEG}
\overline{(H_{\mu \nu})^2} = \sum_\alpha (D_{\mu
  \nu}(\alpha))^2 \ .
\ee
The properties of the matrices $D_{\mu \nu}$ are obtained by counting.
Let $\mu$ and $\nu$ differ in the occupation numbers of $f$
single--particle states. (a) If $f \geq 3$, then $\sum_\alpha (D_{\mu
\nu}(\alpha))^2 = 0$. (b) If $f = 2$, then $\sum_\alpha (D_{\mu
\nu}(\alpha))^2 = 1$. (c) If $f = 1$, then $\sum_\alpha (D_{\mu
\nu}(\alpha))^2 = (m - 1)$. (d) If $f = 0$ or $\mu = \nu$, then
$\sum_\alpha (D_{\mu \mu}(\alpha))^2 = {m \choose 2}$. For the number
of times each of these alternatives is realized, we find (a) $\sum_{j
\geq 3} N {m \choose j} {l-m \choose j}$, (b) $N {m \choose 2} {l - m
\choose 2}$, (c) $N m (l - m)$, and (d) $N$. The number of zero
matrix elements dominates (it is close to $N^2$). The number of unit
values comes next and is approximately $(1/4) N m^2 (l - m)^2$. The
number of values $(m - 1)$ is $N m (l - m)$. The number of values
${m \choose 2}$ is trivially equal to $N$, the dimension of the
matrix and the number of diagonal elements. The correlators
$\sum_\alpha D_{\mu \nu}(\alpha) D_{\rho \sigma}(\alpha)$ can also be
worked out easily but are not given here. They do not all vanish.
Therefore, the random variables $V_\alpha D_{\mu \nu}(\alpha)$ are
not independent. This reflects the fact that the same matrix element
of the two--body interaction may couple different pairs of Slater
determinants.

The individual matrices $D_{\mu \nu}(\alpha)$ have a very simple
structure and barely mix many-body states. Let the TBME corresponding
to index $\alpha$ change the occupation of $f$ particles. For $f=0$ the
matrix $D_{\mu \nu}(\alpha)$ is diagonal. For $f=1,2$ the matrix
$D_{\mu \nu}(\alpha)$ can be ordered to have ${l-2-f\choose m-2}$
block matrices of dimension two and is zero otherwise. This shows that
individual matrices $D_{\mu \nu}(\alpha)$ cannot generate mixing and
chaos.

We consider the limit of large matrix dimension $N$, attained by
taking the limit $l \to \infty$. This can be done in two ways: (i) by
keeping the ratio $m/l$ fixed; (ii) by keeping $m$ fixed. (i) The
ratio of the number of non--diagonal elements with variance unity to
the total number of matrix elements is $(1/4) m^2 (l - m)^2 / N$. This
ratio tends to zero as $l \to \infty$. The same ratio calculated for
the non--diagonal elements with variance $(m - 1)$ vanishes even
faster. The correlators of the diagonal elements are $\sum_\alpha
D_{\mu \mu}(\alpha) D_{\nu \nu}(\alpha) = {k \choose 2}$ where $k$ is
the number of single--particle states that are occupied in both $\mu$
and $\nu$. The correlation is maximal for $k = m - 1$ and vanishes for
$k = 0$ and $k = 1$. For fixed $\mu$ and $k$, the number of states
$\nu$ for which the occupation of $k$ single--particle states is the
same as in $\mu$ is ${l - m \choose m - k} {m \choose k}$. For $k = m
- 1$, this yields $m(l - m)$ and for $k=1$, we find $m {l - m \choose
  m - 1}$. The ratio of both expressions to $N$ vanishes exponentially
fast as $l \to \infty$. Thus, the matrices approach diagonal form with
uncorrelated diagonal elements that all have the same variance. This
is suggestive of the Poissonian distribution. (ii) The ratio of the
number of non--diagonal elements with variance unity to the total
number of matrix elements is $(1/4) m^2 l^2 / N$ and vanishes
exponentially fast. The same holds true {\it a fortiori} for the
elements with variance $m - 1$. Again, the correlators of the diagonal
elements are given by ${k \choose 2}$, and the number of states which
correlate to a given one with this correlator is ${l - m \choose m -
  k} {m \choose k}$. The ratio of this value to $N$ vanishes as a
power of $l$. Again, this is suggestive of the Poissonian
distribution. We note, however, that sparseness of a random matrix is
no guarantee for Poissonian statistics. Indeed, Fyodorov and
Mirlin~\cite{FyoMir91} have shown that sparse random matrices with
${\it uncorrelated}$ non--vanishing matrix elements that have a
frequency $p/N$ and no preference for large diagonal elements may
have either Poissonian or GOE statistics, depending on the value of
$p$. While these results are quite suggestive, the question about
spectral fluctuations of the EGOE(2) in the limit of large matrix
dimension remains open. 

We apply these considerations to the half--filled $sd$-shell
(disregarding, of course, all conserved quantum numbers). The
number of single--particle states is $l = 24$; the number of nucleons
is $m = 12$. The number of independent TBME is $\approx 3.8 \times
10^4$. The corresponding matrices of the embedded ensemble have
dimension $N \approx 2.7\times 10^6$.  The variance of the diagonal
elements is $66$. The fraction of non--diagonal elements with variance
unity is approximately $1.9\times 10^{-3}$, that of those with
variance $m-1 = 11$ is approximately $5.3 \times 10^{-5}$. The
Hamiltonian matrices of the EGOE(2) are thus characterized by their
sparsity, strong diagonal structure, and a relatively large number of
independent TBME.

This structure is very different indeed from that of the GOE. In
analogy to Eq.~(\ref{HamEGOE}), the matrix elements of the GOE
Hamiltonian can be written in the form
\be
H^{\rm GOE}_{\mu \nu} = \sum_{j < l = 1}^N V_{j l}
D^{\rm GOE}_{\mu \nu}(j,l) \ .
\label{HamGOE}
\ee
The random variables $V_{j l}$ are defined for $1 \leq j < l \leq N$
where $N$ is the dimension of $H^{\rm GOE}$ and are uncorrelated
Gaussian--distributed random variables with zero mean and a common
second moment. The matrices $D^{\rm GOE}$ again determine the
structure of the ensemble and are given by
\be
D^{\rm GOE}_{\mu \nu}(j,l) = \delta_{j \mu} \delta_{l \nu} +
\delta_{j \nu} \delta_{l \mu} \ .
\label{DGOE}
\ee
Each such matrix is symmetric and has only one non--vanishing matrix
element above or in the main diagonal. Again, all ensemble averages
of the GOE are determined by the matrices $D^{\rm GOE}$. In view of
the extreme simplicity of these matrices, one usually does not write
the GOE in the form of Eq.~(\ref{HamGOE}). This form is, however,
useful for purposes of comparison. We observe that in the GOE, the
number of independent random variables and, thus, of matrices
$D^{\rm GOE}$ is as large as is consistent with the basic symmetry
(invariance under time reversal) of the ensemble. In the EGOE(2) and
for matrices of the same dimension $N$, the number of independent
random variables and, thus, of matrices $D$ is strongly reduced. This
strong reduction is accompanied by a strong increase in the number of
non--vanishing matrix elements of each of the matrices $D$. Although
sparse, the matrices $D$ are much less so than their GOE counterparts.

\section{Presence of symmetries}
\label{pres}

In this section we consider the TBRE, both for a single $j$--shell
and for the $sd$--shell. We first investigate the TBRE for the case
of a single $j$--shell. This is the simplest case and already exhibits
the major difference to the EGOE. Then we consider the more realistic
(and more complex) case of the $sd$--shell where we encounter several
sub--shells and also have to include isospin in our analysis. 

We consider $m$ fermions in a single $j$--shell. (Later, we will
consider the example $m = 6$ and $j = 19/2$). There are $a = j + 1/2$
TBME as two identical particles can have spins $s=0, 2, 4, \ldots,
(j-1)/2$. The corresponding spin--conserving two--body matrix elements
are denoted by $V_\alpha$, with $\alpha = 1, \ldots, a$. The matrix
elements of the $j$-shell Hamiltonian in the space of many--body
states with total spin $J$ and projection $J_z=0$ are 
\be
\label{Hamj}
H_{\mu \nu}^{J} = \sum_{\alpha=1}^a V_\alpha C_{\mu \nu}^{J}(\alpha) \ .  
\ee
Again, the matrices $C_{\mu \nu}^{J}(\alpha)$ transport the information
about the TBME $V_{\alpha}$ into the space of the many--body states.
With the $V_\alpha$ considered as uncorrelated Gaussian--distributed
random variables with zero mean value and a common second moment, all
properties of the TBRE (mean values, higher moments and correlation
functions of both Hamiltonian and Green's functions) are again
completely determined by the matrices $C^J$, in full analogy to the
cases of the GOE and of the EGOE(2). To see how chaos is generated in
the TBRE it is, thus, neccessary to understand the properties and
structure of these matrices.

In contrast to the EGOE(2), the matrices $C^J$ are determined by both,
the spin symmetry and the fermionic nature of the $m$--body system. Each
element is given in terms of sums over products of angular--momentum
coupling coefficients and coefficients of fractional parentage and is,
thus, a rather complex quantity. Therefore, the properties of the
matrices $C_{\mu \nu}^{J}(\alpha)$ cannot be inferred as easily as
those of the matrices $D_{\mu \nu}$ in Eq.~(\ref{HamEG}). Some facts
can be established analytically. For what remains, we rely on numerical
investigations.

For $j \gg m \gg 1$, the number $N(J)$ of $m$--body states of fixed
total spin $J$ is approximately given by
\be
N(J) \approx \frac{3 (2j+1)^m}{4 \pi j^3 m m!} \sqrt{\frac{6
\pi}{m}} (\delta_{J,0} + 2 J) \exp \{ - \frac{3 J^2}{2 m j^2} \} \ .
\label{numb}
\ee
The right--hand side is the leading term in an asymptotic expansion in
inverse powers of $j$ and $m$. This expression is derived in the
Appendix. We observe that for fixed values of $j$ and $m$, the number
of states increases monotonically with $J$ until the Gaussian cutoff
becomes relevant. The maximum spin has the value $m(2j+1-m)/2$.
The cutoff sets in much below this value. For $J$ fixed and below the
cutoff, $N(J) \propto 2^m j^{m-3} / (m^{3/2} m!)$ grows strongly with
$j$. Thus, for $j \gg m$ the dimension of the matrices $C^J$ is very
much larger than their number $a = j + 1/2$. We note that the trend
seen in the comparison between the GOE and the EGOE(2) continues
unabatedly: In comparison with the matrix dimension, the number of
independent random variables is reduced much below the EGOE(2) value.
To achieve complete mixing of the states, this small number must be
compensated by an increased density of the matrix elements of the
matrices $C^J$.

The $a$ matrices $C^J(\alpha)$ can be viewed as matrix representations
of $a$ operators ${\hat C}^J(\alpha)$. As shown in the Appendix, the
latter are given by
\be
{\hat C}^J(\alpha) = {\bf P}(J) X(\alpha) {\bf P}(J)
\label{oper}
\ee
where ${\bf P}(J)$ is the orthonormal projector onto the subspace of
many--body states with fixed total spin $J$. The operators $X(\alpha)$
are scalar two--body operators normalized in such a way that with
${\hat n}$ the number operator, we have $\sum_\alpha X(\alpha) = (1/2)
({\hat n}^2 - {\hat n})$. Using this representation, it is easy to see
(Appendix) that the ${\hat C}^J(\alpha)$'s do not commute: For $\alpha
\neq \beta$, the commutator $[ {\hat C}^J(\alpha), {\hat C}^J(\beta)]$
is a three--body operator projected onto the space of states with spin
$J$. We also observe that by definition of the operators $X$, we have
$\sum_\alpha C^J_{\mu \nu}(\alpha) = \delta_{\mu \nu} (1/2) m (m - 1)$. 

\begin{figure}[t]
%\vskip 0.3cm
\centerline{
\includegraphics[height=70mm]{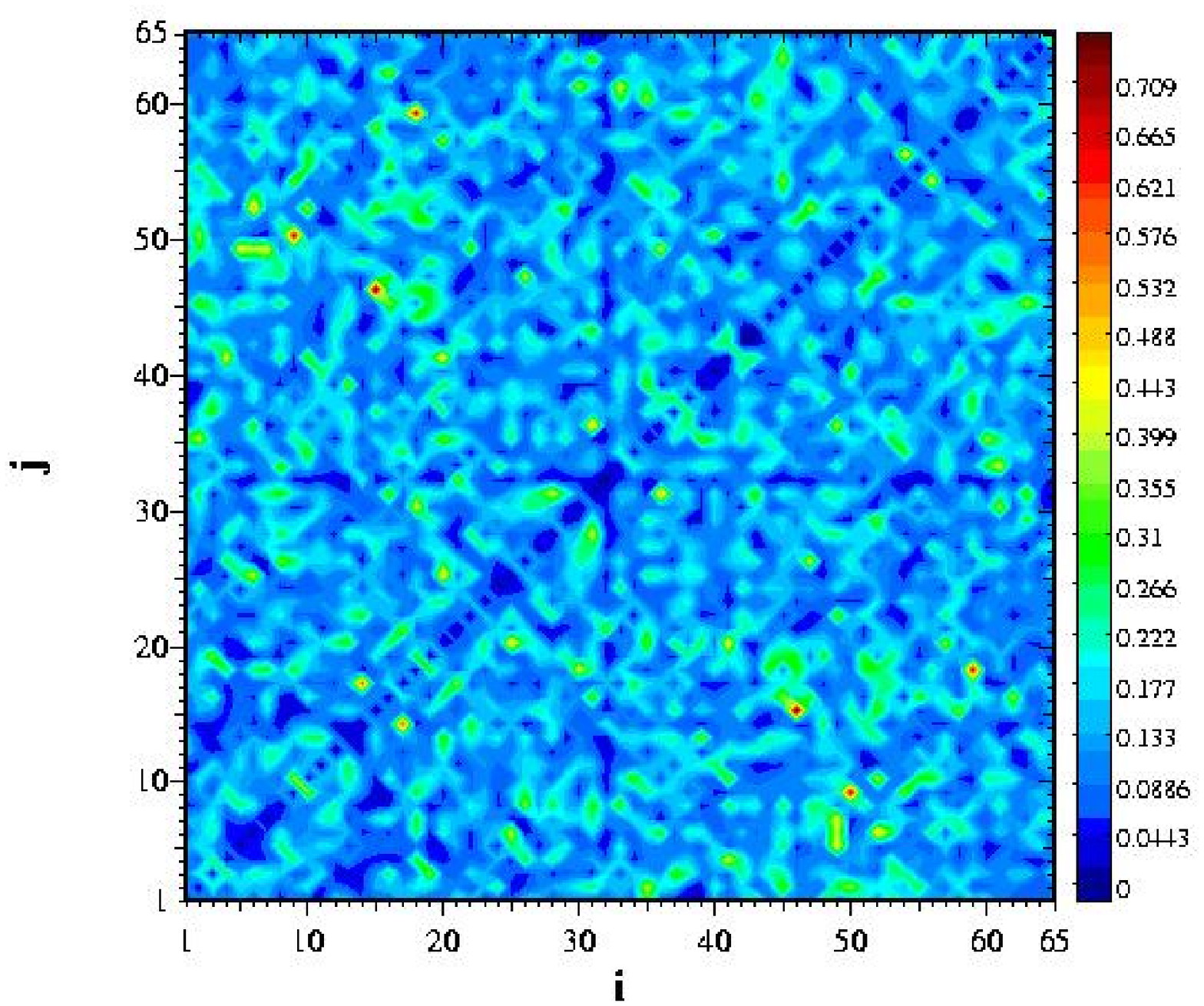}
\includegraphics[height=70mm]{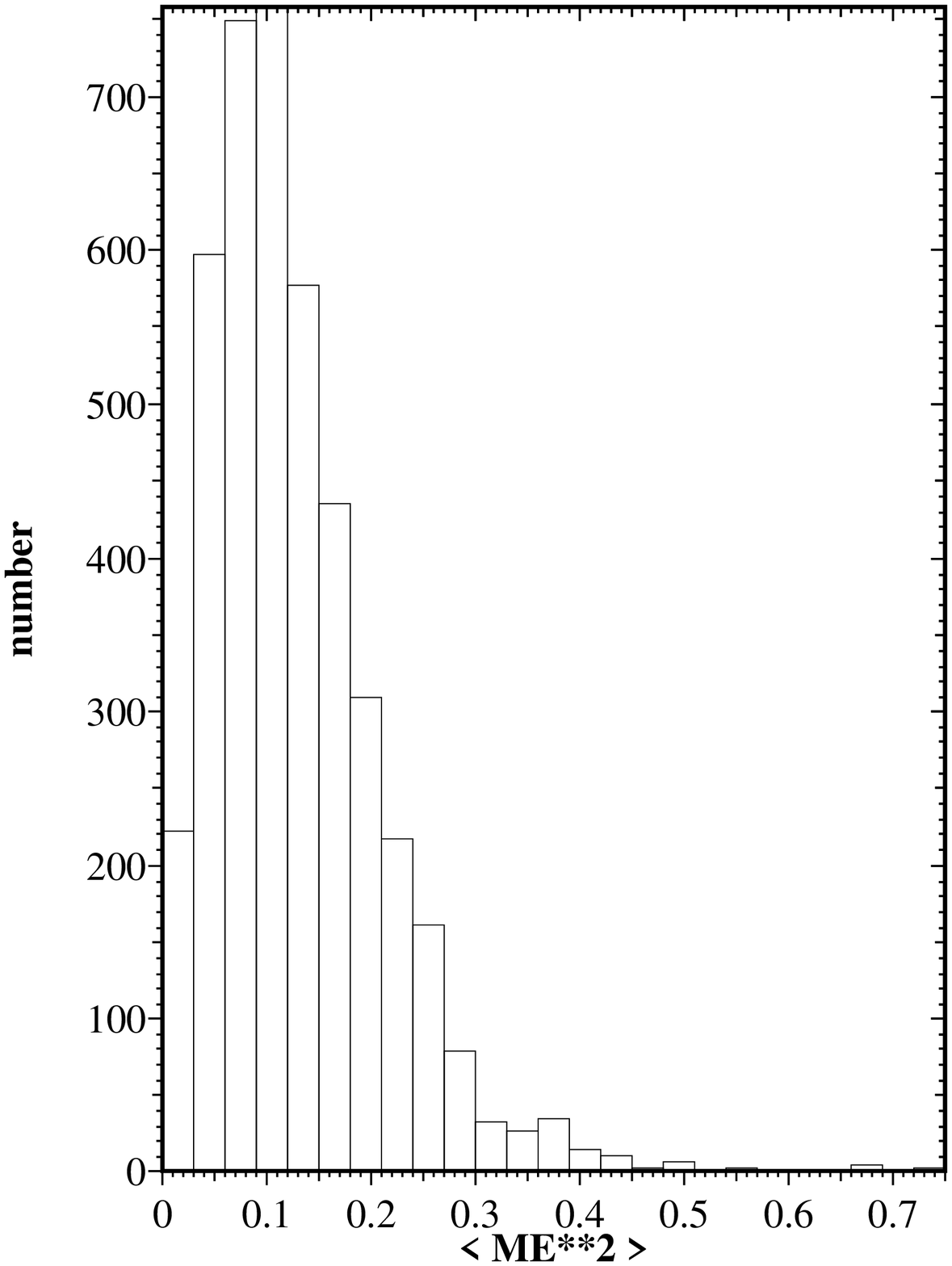}
}
\caption{\label{fig0}Left: ensemble--averaged second moment of the TBRE 
(diagonal elements suppressed) for 6 fermions in a single $j=19/2$ shell
with total spin $J=12$. Right: same data shown in a histogram.
}
\end{figure}

We turn to a numerical determination of the matrices $C^J(\alpha)$ for
a shell with $j = 19/2$ and with $m = 6$ fermions. To this end we have
to define the basis. Total Hilbert space is spanned by Slater
determinants of single--particle states. These have spin projection
$J_z = 0$ (but not well--defined total spin $J$). Basis states with
definite angular momentum are constructed numerically by diagonalizing
the total angular momentum operator $\hat{J}^2$. On the one hand, this
basis is not unique since the spectrum of $\hat{J}^2$ is highly
degenerate. On the other hand, there is no preferred basis, and our
results can therefore be viewed as rather generic. The second moment 
\be
\label{Mom2}
\overline{(H^{J}_{\mu \nu})^2} = \sum_\alpha (C^{J}_{\mu \nu}(\alpha))^2
\ee
exhibits almost constant and dominant diagonal elements, and
considerably smaller off--diagonal elements. The dominance of the
diagonal elements is not surprising in view of the identity $\sum_\alpha
C^{J}_{\mu \nu}(\alpha) = \delta_{\mu \nu} m(m-1)/2$, and in this aspect
the TBRE is similar to the EGOE. The off--diagonal elements of the second
moment (\ref{Mom2}) are depicted in Fig.~\ref{fig0} for the total spin
$J=12$. (This is the largest--dimensional sector in Hilbert space). The
left part of Fig.~\ref{fig0} shows a contour plot of the off--diagonal
elements, while the right part of Fig.~\ref{fig0} shows the corresponding
histogram. Clearly, all off--diagonal elements are non--zero, and this is
in stark contrast to the sparse EGOE matrices. We recall that the
corresponding histogram for the EGOE would have a (giant) peak at zero
and two smaller peaks. This suggests that already a two--body operator
corresponding to a single non--vanishing TBME will strongly mix the basis
states in the $j$--shell TBRE. We recall that the matrices $C_{\mu
\nu}^{J}(\alpha)$ do not commute. Thus, the mixing is expected to be
strong for almost all Hamiltonians of the TBRE. Moreover, the
non--commutativity of the $C^J$'s and the sum rule $\sum_\alpha C_{\mu
\nu}(\alpha) = \delta_{\mu \nu} m(m-1)/2$ (which is invariant under a
rotation of the basis) together strongly suggest that the results shown
in Fig.~\ref{fig0} are generic and independent of the basis chosen.

To lend further substance to these arguments, we have diagonalized the
operators $X(\alpha)$ in the space of the 1242 Slater determinants $|i
\rangle$ with $J_z = 0$ (but with no fixed spin) where $i = 1, \ldots,
D$ with $D = 1242$. The eigenstates carry the labels $\alpha$ and $J$ as
well as a running label $\mu$ where $\mu = 1, \ldots, d_J$ and $d_J$
is the dimension of the subspace with spin $J$. They are given by
$|\alpha J \mu \rangle = \sum_{i = 1}^D c_i(\alpha, J, \mu) | i \rangle$. 
The complexity of the eigenstates $|\alpha J \mu \rangle$ is measured by
the number of principal components ($NPC$) defined as
\be
(NPC)^{-1} = \sum_{i = 1}^D c_i^4(\alpha,J,\mu).
\ee
Table~\ref{tab2} shows the $NPC$s, averaged over all two--body spins
(i.e., over all values of $\alpha$) and over all $d_J$ states with spin
$J$. We recall that the GOE expectation for $NPC$ is $NPC/D = 1/3$. We
see that the spin--conserving two--body operators $X(\alpha)$
individually yield a strong mixing of the basis states. This mixing is
induced entirely by the rotational symmetry and independent of any
particular choice of the two--body interaction. It may be argued that the
complexity of the coefficients $c_i(\alpha,J,\mu)$ reflects just the need
to couple the states $|i \rangle$ to total spin $J$, and is not indicative
of strong mixing due to $X(\alpha)$. Inspection of the individual
coefficients $c_i(\alpha,J,\mu)$ shows, however, strong variations with
the index $\alpha$ which invalidates this argument. Moreover, calculation
of the $NPC$s in a basis with fixed $J$ confirms our picture.

\begin{table}[h]
\begin{tabular}{|r|r|r||r|r|r||r|r|r|}
\hline
 $J$ & $d_J$ & $NPC/D$ &$J$ & $d_J$ & $NPC/D$ &$J$ & $d_J$ & $NPC/D$ \\\hline
 0 & 10 & 0.27 &
 1 & 6 & 0.27 &
 2 & 23 & 0.24 \\
 3 & 21 & 0.23 &
 4 & 37 & 0.24  &
 5 & 32 & 0.24  \\
 6 & 49 & 0.24  &
 7 & 43 & 0.24  & 
 8 & 56 & 0.25  \\
 9 & 51 & 0.25  &
 10 & 62 & 0.25 & 
 11 & 54 & 0.25  \\
 12 & 65 & 0.26  &
 13 & 56 & 0.26  &
 14 & 63 & 0.26  \\
 15 & 55 & 0.26  &
 16 & 60 & 0.26  &
 17 & 50 & 0.26  \\
 18 & 55 & 0.26  &
 19 & 45 & 0.26  &
 20 & 47 & 0.27  \\
 21 & 39 & 0.26  &
 22 & 40 & 0.27  &
 23 & 31 & 0.26  \\
 24 & 33 & 0.26  &
 25 & 25 & 0.26  &
 26 & 25 & 0.26  \\
 27 & 19 & 0.24  &
 28 & 19 & 0.24  &
 29 & 13 & 0.23  \\
 30 & 14 & 0.24 &
 31 & 9 & 0.22 &
 32 & 9 & 0.21 \\
 33 & 6 & 0.21 &
 34 & 6 & 0.21 & 
 35 & 3 & 0.20  \\ 
 36 & 4 & 0.20  &
 37 & 2 & 0.18  &
 38 & 2 & 0.17  \\
 39 & 1 & 0.16  &
 40 & 1 & 0.15  &
 42 & 1 & 0.13  \\\hline
\end{tabular}
\caption\protect{\label{tab2} Number of principal components $NPC$
  normalized by the dimension $D$ for a system of 6 fermions in a
  single $j = 19/2$ shell. The $NPC$'s are averaged over all two--body
  spins and over the $d_J$ states with spin $J$. The GOE expectation
  for infinite matrix dimension is 1/3.} 
\end{table}

We turn to a more complex (and more realistic) shell model consisting
of several sub--shells. This situation is typical for nuclei. We
distribute $m$ valence nucleons over $q$ single--particle subshells
with total angular momenta $j_k, k=1, \ldots, q$. To simplify the
notation, we drop the parity quantum number. The many--body states
labelled $\mu$, $\nu$ have spin and isospin quantum numbers $J$ and
$T$, respectively. The spin--isospin coupled reduced TBME $V^{st}(j_k,
j_l;j_m,j_n)$ (with $k \le l$, $m \le n$, and $s$ and $t$ the
two--body spin and isospin, respectively) are labeled $V_\alpha$ where
$\alpha = 1, \ldots, a$ and where $a$ depends upon the particular
shell under consideration.

The shell--model Hamiltonian has matrix elements 
\be
\label{HamSM}
H_{\mu \nu}^{JT} = \sum_{\alpha=1}^a V_\alpha C_{\mu \nu}^{JT}(\alpha)
\ .  
\ee
Again, the $V_\alpha$'s are considered uncorrelated
Gaussian--distributed random variables with zero mean value and a
common second moment. The matrices $C_{\mu \nu}^{JT}(\alpha)$ contain
the geometric aspects of the shell model. Moreover, all ensemble
averages (moments and correlation functions of the Hamiltonian and of
the Green's functions) of this TBRE depend only upon the $C_{\mu
\nu}^{JT}(\alpha) $'s. Again, it is of central importance to study the
properties and the structure of these matrices. Because of the existence
of subshells, the matrices $C_{\mu \nu}^{JT}(\alpha)$ exhibit a more
complex structure than the EGOE matrices or even the matrices
encountered for a single $j$--shell. To understand the structure of
these matrices, we use first a qualitative analysis. This analysis
applies to the $sd$--shell and to other shells with more than one
subshell in heavier nuclei. We label the $JT$--coupled many--body basis
states by the occupation numbers $(n_1, \ldots, n_q)$ of the $q$
subshells. Here $(n_1, \ldots, n_q)$ is a partition of $m$ into $q$
integers, so that $\sum_{k=1}^q n_k = m$. We assume that the many--body
basis states belonging to a given set of quantum numbers $JT$ are
ordered in blocks, each block containing states belonging to the same
partition. The reduced TBME fall into two classes. The first class
consists of the ``diagonal'' reduced TBME $V^{st}(j_k,j_l;j_k, j_l)$.
These couple only states within the same partition. The matrices
$C_{\mu \nu}^{JT}( \alpha)$ corresponding to these TBME are
block--diagonal. The second class consists of the ``off--diagonal''
reduced TBME $V^{st}(j_k,j_l, j_m,j_n)$ with $(j_k, j_l)\ne (j_m, j_n)$.
These TBME change the occupation numbers $(n_1, \ldots, n_q)$ and, thus,
couple different partitions. Among these off--diagonal TBME there are
those with $k\ne m,n$ and $l\ne m,n$, and those with $k = m$, or $k =
n$, or $l = m$, or $l = n$. The former (latter) change the occupation
numbers of the subshells by two (one) units, respectively. The matrices
$C_{\mu \nu}^{JT}(\alpha)$ corresponding to these off--diagonal reduced
TBME have non--zero entries in the off--diagonal blocks only. As a
result, the matrix $H_{\mu \nu}^{JT}$ attains a checker--board pattern
which reflects the partitions. The pattern is that of $m$ {\it bosons}
distributed over $q$ single--particle orbitals that interact via
two--body interactions.

\begin{figure}[t]
%\vskip 0.3cm
\centerline{
\includegraphics[height=70mm]{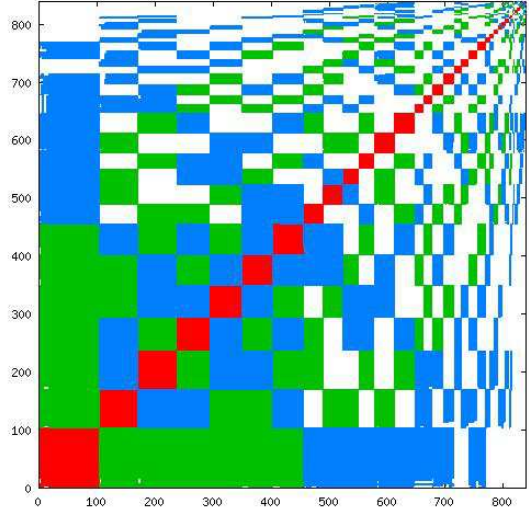}}
\caption{\label{fig1}Non--zero matrix elements of the
$sd$--shell many--body Hamiltonian with $m = 12$ nucleons
(``$^{28}$Si'') in the sector $J = T =0$. Matrix elements originating
from diagonal reduced two--body matrix elements are red. Matrix
elements originating from off--diagonal reduced two--body matrix
elements that transfer one (two) particles between partitions are
green (blue).}
\end{figure}

We display this structure for the particular case of the $sd$--shell
with $l = 24$ single--particle states, $q = 3$ subshells with $j_1 =
5/2$ (for the $d_{5/2}$ sub--shell), $j_2 = 3/2$ (for the $d_{3/2}$
sub--shell, and $j_3 = 1/2$ (for the $s_{1/2}$ sub--shell), occupation
numbers $n_1, n_2, n_3$, and a total of $41$ partitions. Among the $a
= 63$ reduced TBME, 28 are diagonal, 22 of the off--diagonal reduced
TBME induce one--body transitions between the partitions, and 13
transfer two particles. We consider the case of $m = 12$ nucleons
(``$^{28}$Si'') in the sector $J = T = 0$. The resulting matrix has
dimension $839$. All calculations were done using the shell--model
code {\sc Oxbash}~\cite{oxbash}. Figure~\ref{fig1} shows the matrix
structure (non--zero matrix elements only) of the
Hamiltonian~(\ref{HamSM}). The partition structure and the resulting
checker--board pattern are clearly visible. 
 
To elucidate further the structure of the shell--model Hamiltonian
of Eq.~(\ref{HamSM}), we ask: How many matrices $C_{\mu \nu}^{JT}
(\alpha)$ yield a non--zero contribution to a given matrix element
$H_{\mu \nu}^{JT}$? A matrix element of the diagonal blocks gets on
average non--zero contributions from $25.2 \pm 4.0$ of the 28
block--diagonal matrices corresponding to the diagonal reduced TBME.
For the off--diagonal--block matrices corresponding to off--diagonal
reduced TBME, the numbers are $7.2 \pm 1.4$ out of 22 and $2.0 \pm
0.9$ out of 13 for those transferring one and two particles between
different partitions, respectively. This shows that in its non--zero
blocks, each of the individual matrices $C_{\mu \nu}(\alpha)$ is
rather densely populated, and that the density decreases with
increasing number of transferred particles.

Further insight into the structure of the Hamiltonian~(\ref{HamSM}) is
obtained when we confine attention to the block--diagonal matrices
$C_{\mu \nu}(\alpha)$ associated with diagonal TBME. Similar to what
was done for the TBRE for a single $j$--shell, we diagonalize these
matrices and compute the average number of principal components
($NPC$), i.e. the inverse of the sum over the expansion coefficients
raised to the fourth power~\cite{Zelevinsky}, for each partition. The
average is taken over all eigenstates within one partition and over
the ensemble of block--diagonal matrices $C_{\mu \nu}(\alpha)$ and is
shown in Table~\ref{tab1}. We recall that for the GOE and in the limit
of infinite matrix dimension $D$, the value would be $NPC / D = 1 /
3$: a typical GOE eigenstate has significant overlap with about $1 /
3$ of the basis states.  Table~\ref{tab1} shows that $NPC / D$ amounts
to between 25\% and 70\% of the GOE expectation for the partitions
with dimension $D > 20$. The partitions with smaller dimensions yield
even higher values for $NPC/D$, but the fluctuations typically increase
with decreasing dimension. The degree of mixing tends to decrease with
increasing dimension of the partitions although there are considerable
fluctuations. We conclude that each diagonal reduced TBME thoroughly
mixes the states belonging to a fixed partition. To fully appreciate
these statements, the reader should recall that the matrices $C_{\mu
\nu}(\alpha)$ do not carry any specific information on the two--body
interaction and are determined exclusively by the symmetries of the
problem (exclusion principle and conserved quantum numbers $J$ and
$T$). The mixing within each partition is expected to be even stronger
when we consider a generic two--body interaction and the resulting
superposition of matrices $C_{\mu \nu}^{JT}(\alpha)$ in
Eq.~(\ref{HamSM}).

\begin{table}[h]
\begin{tabular}{|c||r|r|}
\hline
Partition  & $D$  & $NPC/D$ \\\hline
$(6,4,2)$ & 103 & 0.10 \\
$(7,3,2)$ &  67 & 0.11 \\
$(5,5,2)$ &  67 & 0.14 \\
$(5,4,3)$ &  56 & 0.18 \\
$(7,4,1)$ &  56 & 0.09 \\
$(6,3,3)$ &  53 & 0.15 \\
$(6,5,1)$ &  53 & 0.13 \\
$(8,2,2)$ &  35 & 0.20 \\
$(4,6,2)$ &  35 & 0.21 \\
$(4,5,3)$ &  34 & 0.23 \\
$(8,3,1)$ &  34 & 0.14 \\
$(7,2,3)$ &  27 & 0.22 \\
$(5,6,1)$ &  27 & 0.24 \\\hline
\end{tabular}
\caption\protect{\label{tab1} Number of principal components $NPC$
  normalized by the dimension $D$ for the 13 largest partitions
  generated by mixing due to the block--diagonal matrices
  $C_{\mu \nu}(\alpha)$. The GOE expectation for infinite matrix
  dimension is 1/3.}
\end{table}

Although the mixing between partitions is not as strong as that
within each partition, $H_{\mu \nu}^{JT}$ will, for $^{28}$Si and $J
= 0$ $T = 0$, generically generate chaos. To show this, we have
compared our Figure~\ref{fig1} with the corresponding figure for the
Wildenthal two--body interaction (which was used in
Ref.~\cite{Zelevinsky} and shown there to produce chaos; see, e.g.,
Figs.~23(a) and 23(c) of that reference). The two figures are
indistinguishable.

\section{Discussion and Summary} 

The aims and results of our paper can be summarized from two points
of view, from that of random--matrix theory and from that of nuclear
structure theory.

From the viewpoint of random--matrix theory, we have compared three
random--matrix ensembles that have been used in the past to deal
with interacting many--body systems and, in particular, with nuclei:
The GOE, the EGOE(2), and the TBRE, the latter applied to a single
$j$--shell and to the $sd$--shell. In our comparison, we have
emphasized the central role of the structure matrices denoted by
$D^{\rm GOE}$, $D$, $C^{J}$, and $C^{JT}$, respectively. These
matrices provide the scaffolding of the underlying random--matrix
ensemble. Averages, higher moments and correlation functions of all
observables can be expressed in terms of and are completely determined
by these matrices. These matrices then must form the central object
of study of the said random--matrix ensembles.

In the framework of a many--body problem, use of the GOE is tantamount
to assuming many--body forces the rank of which equals the number of
valence particles. This unrealistic aspect of the GOE has to be
weighed against the great advantage of its structural simplicity.
This simplicity is due to the orthogonal invariance of the GOE. It
allows for a complete analytical calculation of all moments and
correlation functions. Hence the predictive power of the GOE. The
orthogonal invariance manifests itself in the large number ($\propto
N^2$ with $N$ the matrix dimension) of independent random variables
and in the extreme simplicity of the structural matrices $D^{\rm GOE}$.
In the GOE, no reference is made to the possible existence of quantum
numbers like spin or isospin.

The EGOE(2) is designed to deal with many--body systems which are
governed by two--body forces. In this respect the EGOE(2) is a much
more realistic model than the GOE. The model does not possess the
orthogonal invariance of the GOE. This is why all attempts at
calculating spectral fluctuations analytically for this ensemble have
failed so far. For the same matrix dimension, the number of
uncorrelated random variables is much smaller than in the GOE. To
achieve complete mixing of the basis states, this reduction must be
made up for by a greater complexity of the matrices $D$. We have
displayed the structure of these matrices which individually are quite
sparse but jointly provide for strong mixing of the basis states. Like
the GOE, the EGOE(2) does not allow for the possible existence of
quantum numbers like spin or isospin.

The TBRE is more realistic yet than the EGOE(2) in that it does take
account of the two essential properties of the residual interaction
of the nuclear shell model: The interaction is a two--body interaction,
and it preserves spin, parity, and isospin. The existence of symmetries
associated with these quantum numbers has an essential influence on the
structure of the ensemble and of the matrices $C^{J}$ and $C^{JT}$
which embody this structure: Compared to the EGOE(2) with the same
matrix dimension, the number of independent random variables is much
reduced once again, but the complexity of the matrices $C^{J}$ and
$C^{JT}$ is much increased. It appears that the TBRE is even harder to
deal with than the EGOE(2). In any case, we are not aware of any
previous attempts to study this ensemble analytically. In the TBRE for
a single $j$--shell, the non--commuting matrices $C^{J}$ have strong
diagonal elements (a feature already encountered for the EGOE(2) and
related to sum rules). The non--diagonal elements are not sparse but,
on the contrary, dense. This is how a complete mixing of the basis
states is achieved in the $j$--shell TBRE. In the $sd$--shell TBRE,
the existence of sub--shells and the associated block structure of
the matrices $C^{JT}$ lends greater complexity yet to the ensemble.
Mixing is very strong within the diagonal blocks and weaker for the
off--diagonal ones. In both these ensembles, symmetries play an
important role and, together with the exclusion principle, define the
structure of the matrices $C^{J}$ and $C^{JT}$. It goes without saying
that the relevant symmetry operators must not from a complete set of
commuting operators as otherwise there would be no room left for a
random--matrix ensemble. 

From the point of view of nuclear structure theory, we have uncovered
generic features of shell--model calculations. These are embodied in
the matrices $C^{J}$ and $C^{JT}$. Every shell--model calculation for
a single $j$--shell or for the $sd$--shell amounts to choosing a
specific linear combination of these matrices (the same for every value
of spin $J$), and diagonalizing the resulting Hamiltonian matrix.
Corresponding statements apply for other shells. Inasmuch as there is
evicence for chaos in one such calculation using a specific set of
two--body matrix elements, the properties of the matrices $C^{J}$ and
$C^{JT}$ displayed above guarantee that spectra and eigenfunctions
calculated for most other choices of the two--body interaction will
likewise be chaotic. The two--body interactions for which this
statement does not apply form a set of measure zero.

It is in this sense that we have demonstrated that chaos is a generic
property of the nuclear shell--model. We have also shown that
symmetries (which are due to the existence of an incomplete set of
commuting operators) determine the structure of the TBRE in an
essential way. Thus, such symmetries are vital for the occurrence of
chaos in the TBRE. Another aspect of our work (not emphasized in the
present paper but a natural spin--off) is the existence of correlations
between many--body spectra having different quantum numbers like total
spin $J$. It is immediately obvious from Eqs.~(\ref{Mom2}) and
(\ref{HamSM}) that Hamiltonian matrices pertaining to different spin
values in the same nucleus are correlated since they depend on the same
set of random variables. This fact has been used to explain the observed
preponderance of spin--zero ground states in calculations using the
TBRE~\cite{pap}.

Our considerations are restricted to spherical nuclei and totally
degenerate major shells. We have remarked in the Introduction that
lifting this degeneracy by taking into account the differences of
the single--particle energies in individual sub--shells, will drive
the system toward regularity. Our considerstions do not apply to
deformed nuclei. Here the Nilsson model provides a single--particle
basis in which it is meaningless to assume degenerate single--particle
energies. Such nuclei possess well--developed collective motion.
However, collective motion exists also beyond the regime of
well--deformed nuclei. It has been notoriously difficult in the past
to understand this fact in the framework of the spherical shell model,
with the exception of certain types of collectivity like that of the
giant dipole resonance. Understanding both, collectivity and chaos,
within a common framework is, thus, a goal of future work. We observe,
however, that collectivity typically involves levels with different
quantum numbers (like those forming a rotational band) while chaos
is a property displayed by levels with identical quantum numbers.
Thus, in nuclei chaos and collectivity need not be antagonistic.

\section*{Acknowledgements}
We are grateful to O. Bohigas and U. Smilansky for helpful
discussions. This research was supported in part by the
U.S. Department of Energy under Contract Nos.\ DE-FG02-96ER40963
(University of Tennessee) and DE-AC05-00OR22725 with UT-Battelle, LLC
(Oak Ridge National Laboratory).

\section*{Appendix}

In the framework of the TBRE for a single $j$--shell, we derive
Eq.~(\ref{numb}), and we define the operators $X(\alpha)$. The
calculation of $N(J)$, the number of many--body states with spin $J$,
is rather standard. We first calculate the number $n(M)$ of states with
$J_z = M$ and then $N(J)$ from the identity $N(J) = n(J) - n(J+1)$. We
focus attention on large values of $m$ and $j$, $m \gg 1$ and $j \gg
1$. We have
\begin{equation}
n(M) = \sum_{-j \leq \mu_1 < \mu_2 < \ldots < \mu_m \leq j}
\delta(M - \sum_{i = 1}^m \mu_i) \ .
\label{a1}
\end{equation}
Here the delta symbol stands for a Kronecker delta and not for the
Dirac delta function. We write the sum as
\begin{equation}
n(M) = \frac{1}{m!} \sum_{-j \leq \mu_1, \mu_2, \ldots, \mu_m \leq j}
\bigg( \prod_{k < l} [1 - \delta_{\mu_k \mu_l}] \bigg) \delta(M -
\sum_{i = 1}^m \mu_i) \ .
\label{a2}
\end{equation}
We expand the product in powers of the Kronecker deltas and consider
first the term of zeroth order $n_0(M)$, i.e., the term with only one
constraint ($M = \sum \mu_i$). It is
\begin{equation}
n_0(M) = \frac{1}{m!} \sum_{\mu_1 = - j}^j \sum_{\mu_2 = - j}^j
\times \ldots \times \sum_{\mu_m = - j}^j \delta(M - \sum_{i = 1}^m
\mu_i) \ . 
\label{a3}
\end{equation}
The Kronecker delta is written as $\delta_{l,0} = (1 / (2 \pi))
\int_{- \pi}^{+ \pi} \exp [ i l \phi ] {\rm d}\phi$. This yields
\begin{equation}
n_0(M) = \frac{1}{2 \pi m!} \int_{- \pi}^{+ \pi} {\rm d}\phi
\sum_{\mu_1, \mu_2, \ldots, \mu_m = - j}^j \exp [ i \phi (M - \mu_1
- \mu_2 - \ldots - \mu_m) ] \ .
\label{a4}
\end{equation}
The summation yields
\be
n_0(M) = \frac{1}{2 \pi m!} \int_{- \pi}^{+ \pi} {\rm d}\phi \
\cos [ \phi M ] \biggl( \frac{\sin [ (j + 1/2) \phi ]}{\sin \phi
/ 2} \biggr)^m \ .
\label{a5}
\ee
We replace $\phi$ by $x = j \phi$ and use that $j \gg 1$. Then the
function
\be
\bigg( \frac{ \sin [ (1 + 1/(2j)) x ]}{ \sin [ x / (2 j) ]} \bigg)^m
\label{a7}
\ee
has maxima at $x = 0$, $x = \pm 3 \pi/2$, $x = \pm 5 \pi / 2$,
$\ldots$ with absolute values $(2j)^m$, $(2j)^m (2/(3 \pi))^m$,
$(2j)^m (2 / (5 \pi))^m$, $\ldots$. For $m \gg 1$, the maximum
at $x = 0$ gives the dominating contribution. We take account of
this maximum only and write
\begin{eqnarray}
\biggl( \frac{\sin [ (1 + 1/(2j)) x ]} {\sin [ x / (2 j) ]}
\biggr)^m &=& \exp \bigg\{ m \ln \sin [(1 + 1/(2j))x] - m \ln \sin
[x / (2j)] \bigg\} \nonumber \\
&& \qquad \approx (2j+1)^m \exp \{ - (m/6) x^2 \} \ . 
\label{a8}
\end{eqnarray}
In the exponent we have omitted terms of higher order than the first
in $x^2$. This is justified because upon using a Taylor expansion and
performing the Gaussian integration, such terms will produce inverse
powers of $m$ and are, therefore, negligible. Thus, our procedure
amounts to an asymptotic expansion in inverse powers of $j$ and $m$.
Hence,
\begin{eqnarray}
n_0(M) &\approx& \frac{(2j+1)^m}{2 \pi j m!} \int_{- \infty}^{\infty}
{\rm d} x \ \cos [ M x / j ] \exp \{ - (m/6) x^2 \} \nonumber \\
&=& \frac{(2j+1)^m}{2 \pi j m!} \sqrt{ \frac{6 \pi}{m}} \exp \big\{ -
\frac{3M^2}{2mj^2} \big\} \ .
\label{a9}
\end{eqnarray}

We turn to the terms which are linear in the Kronecker delta's in
Eq.~(\ref{a2}). Their sum is denoted by $n_1(M)$ and can be calculated
along quite similar lines. In the same asymptotic limit we find
\be
n_1(M) \approx - \frac{m^2 n_0(M)}{2(2j+1)} \ .
\label{a10}
\ee
This shows that $n_1(M)$ is negligible in comparison with $n_0(M)$ if
$j \gg m$. That same statement holds {\it a fortiori} for the
contributions which are of higher order in the Kronecker delta's. Thus,
$n(M) \approx n_0(M)$, and $N(J) = n(J) - n(J+1)$ then gives the
result~(\ref{numb}).

We come to the definition of the operators $X(\alpha)$. Let $a^{}_\rho$
and $a^{\dagger}_\rho$ be the destruction and creation operators for a
fermion in a state with $j_z$--component $\rho$ with $\rho = -j, \ldots,
j$. When acting upon the vacuum state, the operators
\be
(A^s_M)^{\dagger} = \sum_\rho c(j j s; \rho ,M-\rho) a^{\dagger}_\rho
a^{\dagger}_{M-\rho}
\label{b1}
\ee
create a pair of fermions coupled to total spin $s$ and $z$--component
$M$. Here, $s = 0,2,4,\ldots,2j-1$. The Hermitean conjugate operators
are
\be
A^s_M = - \sum_\rho c(j j s; \rho ,M-\rho) a^{}_\rho a^{}_{M-\rho} \ .
\label{b2}
\ee
The scalar operators
\be
X(s) = (1/2) \sum_M (A^s_M)^{\dagger} A^s_M
\label{b3}
\ee
describe the interaction of two fermions coupled to spin $s$. With
$\alpha = (1/2)s + 1$, these are the operators $X(\alpha)$ introduced
in Eq.~(\ref{oper}). From an identity for the Clebsch--Gordan
coefficients, it follows trivially that
\be
\sum_s X(s) = (1/2) {\hat n} ({\hat n} - 1) \ . 
\label{b4}
\ee
This is the relation used below Eq.~(\ref{oper}). For $s \neq t$, the
commutator $[ X(s), X(t)]$ does not vanish and has the from of a
three--body interaction term. Since the projection operators ${\bf P}
(J)$ trivially commute with $X(s)$ for all $J$ and $s$, the commutator
of two operators ${\hat C}$ with $\alpha \neq \beta$ is
\be
[{\hat C}^J(\alpha), {\hat C}^J(\beta)] = {\bf P}(J) [ X(\alpha),
X(\beta)] {\bf P}(J)
\label{b5}
\ee
and it follows that the ${\hat C}$'s likewise do not commute.

\end{document}